\documentclass[a4paper]{article}

\usepackage{18lomcon}        
\usepackage{cite}            
\usepackage{epsfig}          
\usepackage{epstopdf}

\usepackage{amsmath,amsfonts,amssymb,amsthm}
\usepackage{float}

\begin{document}


\title{EFFECTS OF BSMS ON $\theta_{23}$ DETERMINATION}

\author{C.R.~Das \email{das@theor.jinr.ru}}
\affiliation{Bogoliubov Laboratory of Theoretical Physics, Joint Institute for Nuclear Research,\\ Joliot-Curie 6, 141980 Dubna, Moscow region, Russian Federation}

\author{Jukka Maalampi \email{jukka.maalampi@jyu.fi}}
\affiliation{University of Jyv\"askyl\"a, Department of Physics, P.O.~Box 35,\\ FI-40014 University of Jyv\"askyl\"a, Finland}

\author{Jo\~ao Pulido \email{pulido@cftp.ist.utl.pt}}
\affiliation{Centro de F\'isica Te\'orica das Part\'iculas (CFTP), Instituto Superior T\'ecnico,\\ Av.~Rovisco Pais, P-1049-001 Lisboa, Portugal}

\author{Sampsa Vihonen \email{sampsa.p.vihonen@student.jyu.fi}}
\affiliation{University of Jyv\"askyl\"a, Department of Physics, P.O.~Box 35,\\ FI-40014 University of Jyv\"askyl\"a, Finland}


\date{}
\maketitle


\begin{abstract}
We investigate the prospects for determining the octant of $\theta_{23}$ in the future long baseline oscillation experiments. We present our results as contour plots on the ($\theta_{23}-45^\circ$, $\delta$)--plane, where $\delta$ is the CP phase, showing the true values of $\theta_{23}$ for which the octant can be experimentally determined at 3$\,\sigma$, 2$\,\sigma$ and 1$\,\sigma$ confidence level, in particular, the impact of the non-unitarity of neutrino mixing.
\end{abstract}

\section{The causatum}
The recent data indicate that the neutrino mixing angle $\theta_{23}$ deviates from the maximal-mixing value of 45$^\circ$, showing two nearly degenerate solutions, one in the lower octant (LO) ($\theta_{23}<45^\circ$) and one in the higher octant (HO) ($\theta_{23}>45^\circ$). This can hamper the interpretation of neutrino oscillation data.

We have presented the sensitivity to the determination of the $\theta_{23}$ octant ($\theta_{23}\leq\pi/4$ or $\theta_{23}\geq\pi/4$) in DUNE in four different scenarios. On the one hand, we have updated the 1$\,\sigma$, 2$\,\sigma$ and 3$\,\sigma$ confidence level contours for the Standard Model, where oscillations are constituted between three active neutrinos. On the other hand, we have also given these contours for three different scenarios where the octant sensitivity is interfered by sterile neutrinos and other potential sources for physics beyond Standard Model. We analyzed these scenarios by parametrizing the new physics with the methods that were originally introduced in Refs.~\cite{Escrihuela:2015wra} and \cite{Blennow:2016jkn} to describe non-unitarity of the light neutrino mixing matrix.

We found that the non-unitarity of the mixing matrix caused the sensitivity $\theta_{23}$ octant to decrease from the Standard Model case. Nevertheless, due to the strictness of the existing bounds for the non-unitarity parameters $\alpha_{i j}$, $i,j=1,2,3$ derived in Ref.~\cite{Escrihuela:2016ube} and for $\alpha_{\ell\ell'}$, $\ell,\ell'=e,\mu,\tau$ derived in Ref.~\cite{Blennow:2016jkn} the observed drop in the octant sensitivity was found to be very small and the bounds are within 90\% and 2$\,\sigma$ confidence levels. The worsening of the octant sensitivity due to sterile neutrino was found larger than this. The sensitivity was calculated in this case using the bounds on $\alpha_{\ell\ell'}$ given in Ref.~\cite{Blennow:2016jkn}. The worsening of the sensitivity was found to be less than 1$^\circ$ in each octant.

\vspace{-0.4cm}
\begin{table}[H]
\begin{center}
\begin{tabular}{|c|c|c|}\hline
\rule{0pt}{3ex}Parameter (95\% CL)& $\Delta m_{41}^2\sim 0.1-1$ eV$^2$ & $\Delta m_{41}^2\geq 100$ eV$^2$\\ \hline
\rule{0pt}{3ex}$\alpha_{e e}$ & $1.0\times 10^{-2}$ & $2.4\times 10^{-2}$\\
\rule{0pt}{3ex}$\alpha_{\mu\mu}$ & $1.4\times 10^{-2}$ & $2.2\times 10^{-2}$\\
\rule{0pt}{3ex}$\alpha_{\tau\tau}$ & $1.0\times 10^{-1}$ & $1.0\times 10^{-1}$\\
\rule{0pt}{3ex}$|\alpha_{\mu e}|$ & $1.7\times 10^{-2}$ & $2.5\times 10^{-2}$\\
\rule{0pt}{3ex}$|\alpha_{\tau e}|$ & $4.5\times 10^{-2}$ & $6.9\times 10^{-2}$\\
\rule{0pt}{3ex}$|\alpha_{\tau\mu}|$ & $5.3\times 10^{-2}$ & $1.2\times 10^{-2}$\\ \hline
\end{tabular}
\end{center}
\vspace{-0.2cm}
\caption{\label{bounds:3}Bounds on non-unitary parameters in $\alpha_{\ell\ell'}$ representation, taken from \cite{Blennow:2016jkn}. In this scenario the constraints would correspond to mixing with a light sterile neutrino in two mass scales: $\Delta m_{41}^2\sim 0.1-1$ eV$^2$ (left column) and $\Delta m_{41}^2\geq 100$ eV$^2$ (right column).}
\end{table}

\vspace{-0.8cm}
\begin{figure}[H]
\begin{center}
\vspace{-0.45cm}
\includegraphics[width=10.48cm]{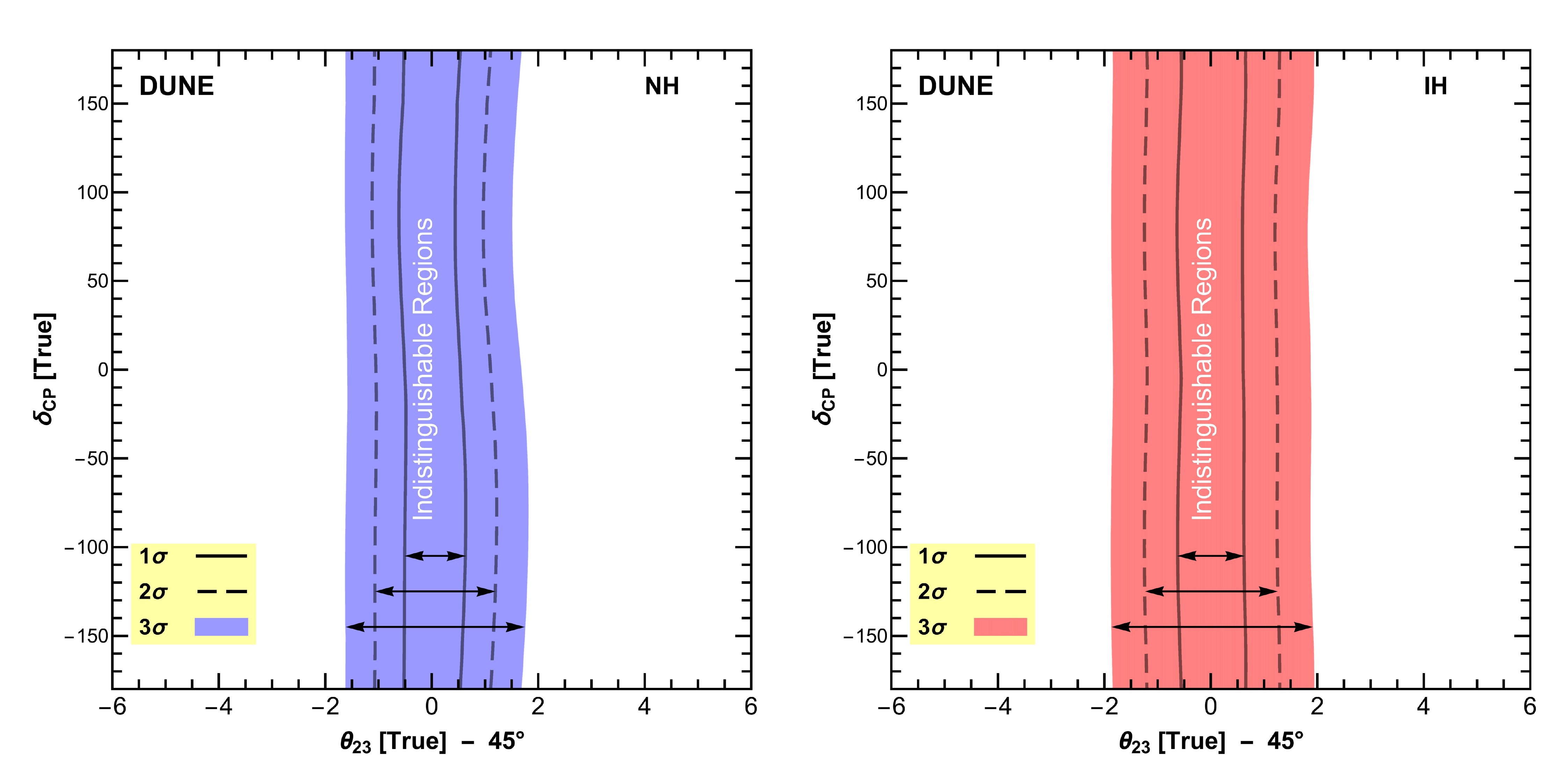}
\vspace{-0.45cm}
\includegraphics[width=10.48cm]{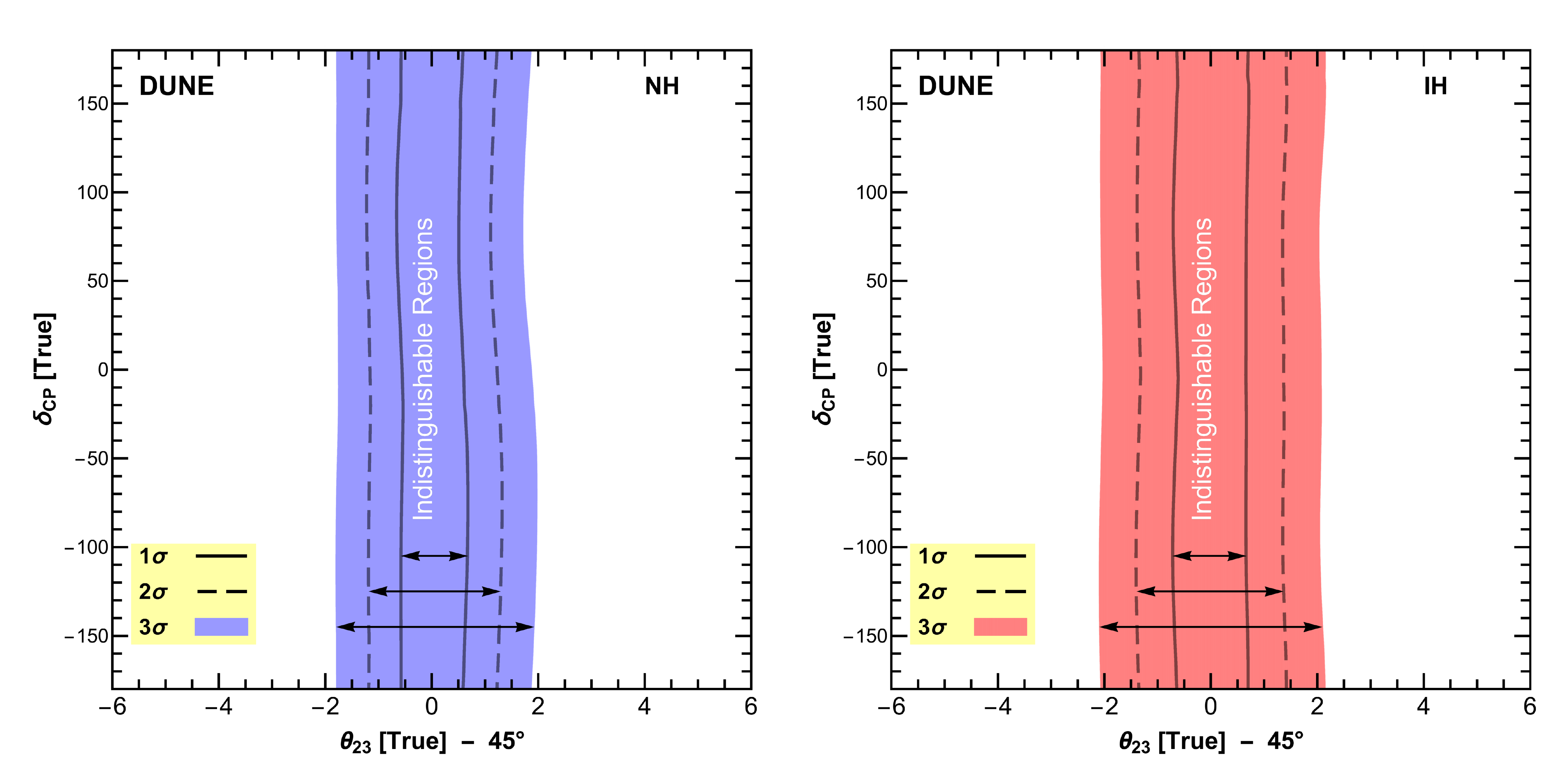}
\caption{\label{fig:1} Octant determination in DUNE in the presence of non-unitary mixing in top two and light sterile mixing with $0<\Delta m^2_{41}<1$ eV$^2$ or $\Delta m_{41}\geq 100$ eV$^2$ in bottom two figures.}
\end{center}
\end{figure}

We found the decrease in sensitivity due to the light sterile neutrino to be substantially less significant than that reported in Ref.~\cite{Agarwalla:2016xlg} where the impact of a sterile neutrino with mixing angles $\theta_{14}=\theta_{24}=9^\circ$ and $\theta_{34}=0^\circ$ was considered in the determination of the $\theta_{23}$ octant in DUNE. Evidence of this sensitivity decrease, can be seen from the comparison between our Figure \ref{fig:1} with Fig.~3 of Ref.~\cite{Agarwalla:2016xlg}. When converted to the non-unitarity formalism (see the appendix of Ref.~\cite{Escrihuela:2015wra}), this kind of sterile neutrino would imply non-unitarity whose parameter values lie close to the existing bounds we presented for $0<\Delta m^2_{41}<1$ eV$^2$ in Table \ref{bounds:3}. On the other hand, our investigation takes into account all possibilities for light sterile neutrinos, whereas the authors of Ref.~\cite{Agarwalla:2016xlg} consider a specific model. Thus our results are in this respect more general, therefore statistically favoured by comparison and hence the difference between the two sets. If the model of Ref.~\cite{Agarwalla:2016xlg} (disfavoured by roughly 95\% CL) is realized in nature, then the ability of DUNE to tell the $\theta_{23}$ octancy is deteriorated.

We also tested how the octant sensitivity changed when the new physics parameters $\alpha_{i j}$ were left unconstrained. This type of simulation corresponds to a new physics scenario, where sterile neutrinos are associated with other new physics effects, not taken into account in Refs.~\cite{Blennow:2016jkn} and \cite{Escrihuela:2016ube} when deriving the bounds for the non-unitary and light sterile mixing effects. An example of this could be non-standard interactions involved in the neutrino propagation. Our simulations showed that in the worst case the octant could be determined at 3$\,\sigma$ CL or better for $\theta_{23}\lesssim 41.0^\circ$ and $\theta_{23}\gtrsim 48.5^\circ$ for the normal hierarchy to be compared with the bounds $\theta_{23}\lesssim 43.5^\circ$ and $\theta_{23}\gtrsim 46.5^\circ$ of the standard case.

\section{The succinct}
We found that non-unitarity of the neutrino mixing matrix or the possible existence of light sterile neutrinos affect only mildly the sensitivity of DUNE to determine the octant of $\theta_{23}$. This is in contrast with the determination of the CP violation, where the presence of sterile neutrinos could jeopardize the sensitivity \cite{Ge:2016xya,Escrihuela:2016ube,Blennow:2016jkn}. See arXiv:1708.05182 for detail analysis.


\end{document}